# Improved Intranuclear Cascade Models for the Codes CEM2k and LAQGSM

S. G. Mashnik*, K. K. Gudima†, A. J. Sierk* and R. E. Prael*

*Los Alamos National Laboratory, Los Alamos, NM 87545, USA
†Institute of Applied Physics, Academy of Science of Moldova, Chişinău, MD-2028, Moldova

**Abstract.** An improved version of the Cascade-Exciton Model (CEM) of nuclear reactions implemented in the codes CEM2k and the Los Alamos version of the Quark-Gluon String Model (LAQGSM) has been developed recently at LANL to describe reactions induced by particles and nuclei at energies up to hundreds of GeV/nucleon for a number of applications. We present several improvements to the intranuclear cascade models used in CEM2k and LAQGSM developed recently to better describe the physics of nuclear reactions. First, we incorporate the photonuclear mode from CEM2k into LAQGSM to allow it to describe photonuclear reactions, not previously modeled there. Then, we develop new approximations to describe more accurately experimental elementary energy and angular distributions of secondary particles from hadron-hadron and photon-hadron interactions using available data and approximations published by other authors. Finally, to consider reactions involving very highly excited nuclei ($E^* \geq 2 - 3$ MeV/A), we have incorporated into CEM2k and LAQGSM the Statistical Multifragmentation Model (SMM), as a possible reaction mechanism occurring after the preequilibrium stage. A number of other refinements to our codes developed recently are also listed.

## INTRODUCTION

Recently, we have developed at Los Alamos National Laboratory an improved version of the Cascade-Exciton Model (CEM) of nuclear reactions contained in the codes CEM2k [1] and the Los Alamos version of the Quark-Gluon String Model (LAQGSM) [2] to describe reactions induced by particles and nuclei at energies up to hundreds of GeV/nucleon for a number of applications. To describe fission and production of light fragments heavier than $^4$He, we have merged our codes with several evaporation/fission/fragmentation models, including the Generalized Evaporation/fission Model code GEM2 by Furihata [3]. CEM2k+GEM2 and LAQGSM+GEM2 perform as well as and often better than other current models in describing a large variety of spallation, fission, and fragmentation reactions. The status of our codes as of the middle of 2003 together with illustrative results and comparisons with other models can be found in [4, 5] and references therein. Here, we present some additional improvements developed during the last year.

It is well known that all IntraNuclear Cascade models (INC) have some problems in a correct description of the experimental spectra of forward-emitted nucleons. Recently, this problem was addressed and partially solved, for example, in the Liege code INCL3 (and INCL4) by Cugnon *et al.* [6] and in the Bruyères-le-Châtel INC code BRIC1.4 by Duarte [7].

Addressing this problem and improving our INC for the CEM2k and LAQGSM codes is a primary aim of this work.

## RESULTS AND DISCUSSION

First, we have incorporated into LAQGSM the improved approximations for the total elastic and inelastic cross sections of hadron-hadron and photo-hadron elementary interactions developed previously for the code CEM2k (see details in [8]).

Second, the photonuclear mode from CEM2k was incorporated into LAQGSM, adding the capability to describe photonuclear reactions at energies from about 30 MeV to about 1.5 GeV (all previous versions of LAQGSM could not model photo-nuclear reactions).

Third, the double differential distributions of secondary particles from elementary $NN$ and $\gamma N$ interactions were simulated by CEM2k (and all its precursors, as well as by LAQGSM and its precursors at energies below 4.5 GeV/A) still using the old Dubna INC [9] approximations that were obtained by Gudima *et al.* [10] 36 years ago, using the measurements available at that time. For instance, in the case of two-body reactions, the cosine of the angle of emission of secondary particles in the c.m. system is calculated by the Dubna INC as a function of a random number $\xi$, distributed uniformly in

the interval [0,1] as

$$\cos\theta = 2\xi^{1/2}\left[\sum_{n=0}^{N} a_n \xi^n + (1 - \sum_{n=0}^{N} a_n)\xi^{N+1}\right] - 1, \quad (1)$$

where $N = M = 3$,

$$a_n = \sum_{k=0}^{M} a_{nk} T_i^k. \quad (2)$$

The coefficients $a_{nk}$ were fitted to the then available experimental data at a number of incident kinetic energies $T_i$, then interpolated and extrapolated to other energies (see details in [9, 10] and references therein). The distribution of secondary particles over the azimuthal angle $\varphi$ is assumed isotropic. For elementary interactions with more than two particles in the final state, the Dubna INC uses the statistical model to simulate the angles and energies of products (see details in [9]).

For the improved versions of our codes referred to as CEM03 and LAQGSM03, respectively, we use currently available experimental data and recently published systematics proposed by other authors to develop new approximations for angular and energy distributions of particles produced in nucleon-nucleon and photon-proton interactions. So, for $pp$, $np$, and $nn$ interactions at energies up to 2 GeV, we did not have to develop our own approximations analogous to the ones described by Eqs. (1) and (2), since reliable systematics have been developed recently by Cugnon et al. for the Liege INC [6], then improved still further by Duarte for the BRIC code [7]: we simply incorporate into CEM03 and LAQGSM03 the systematics by Duarte [7]. Similarly, for $\gamma N$ interactions, we took advantage of the event generators for $\gamma p$ and $\gamma n$ reactions from the Moscow INC [11] kindly sent us by Dr. Igor Pshenichnov. In our codes, we use part of a large data file with smooth approximations through presently available experimental data from the Moscow INC [11] and have developed a simple and fast algorithm to simulate unambiguously $d\sigma/d\Omega$ and to choose the corresponding value of $\Theta$ for any $E_\gamma$, using a single random number $\xi$ uniformly distributed in the interval [0,1] [12]. For other elementary interactions, we fit new sets of parameters $a_n$ from Eq. (1) at different $T_i$ for which we found data, then we approximated energy dependences of the parameters $a_{nk}$ in Eq. (2) using the fitted values of $a_n$.

Two examples of angular distributions of secondary particles from $np$ and $\gamma p$ reactions at several energies are shown in Figs. 1 and 2, respectively. The new approximations from CEM03 and LAQGSM03 reproduce the experimental data much better than the old Dubna INC used in our previous code versions (and in several other codes developed from the Dubna INC).

Additionally, in CEM03 and LAQGSM03 we improved the description of complex-particle spectra. This was done by refining the coalescence model used in our codes, by developing a better approach to estimate the probability of complex-particle emission at the preequilibrium stage of a reaction, and by incorporating into our codes the known systematics for angular distributions of complex particles developed by Kalbach. We do not show these results here since they will be presented in a future paper. Finally, we studied and refined the treatments of the Pauli principle, refraction and reflection from the nuclear potential, details of the nuclear density and the momentum distribution of intranuclear nucleons. Several small uncertainties and observed errors in previous versions of our codes were fixed, leading to the new versions CEM03 and LAQGSM03.

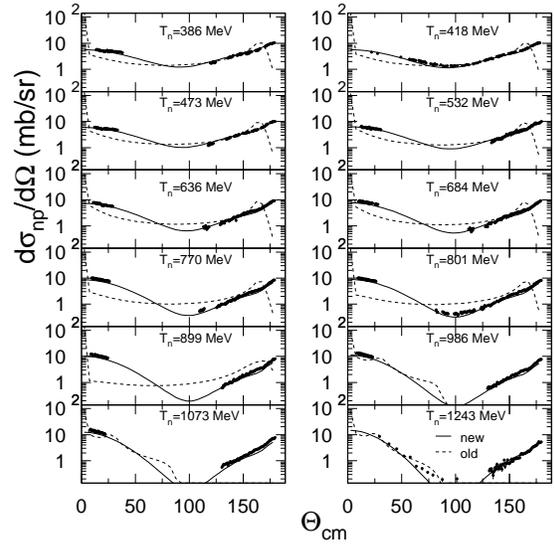

**FIGURE 1.** Example of twelve angular distributions of $n$ from $np$ elastic interactions as functions of $\Theta_{c.m.}^n$ at $T_n$ from 386 to 1243 MeV. The dashed lines show the old approximations from the Dubna INC while the solid lines are the new approximations incorporated into CEM03 and LAQGSM03. Experimental data (symbols) are from [13].

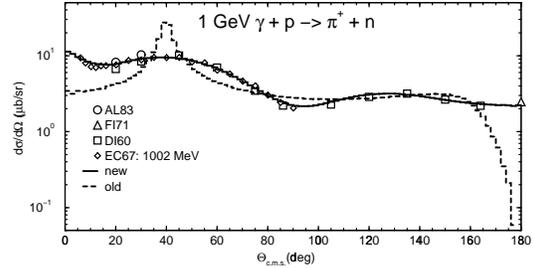

**FIGURE 2.** Angular distribution of $\pi^+$ from 1 GeV $\gamma + p \to \pi^+ + n$ as a function of $\Theta_{c.m.}^\pi$. The dashed line shows the old approximation from the Dubna INC while the solid line is the new approximation incorporated into CEM03 and LAQGSM03. Experimental data (symbols) are from: [14].

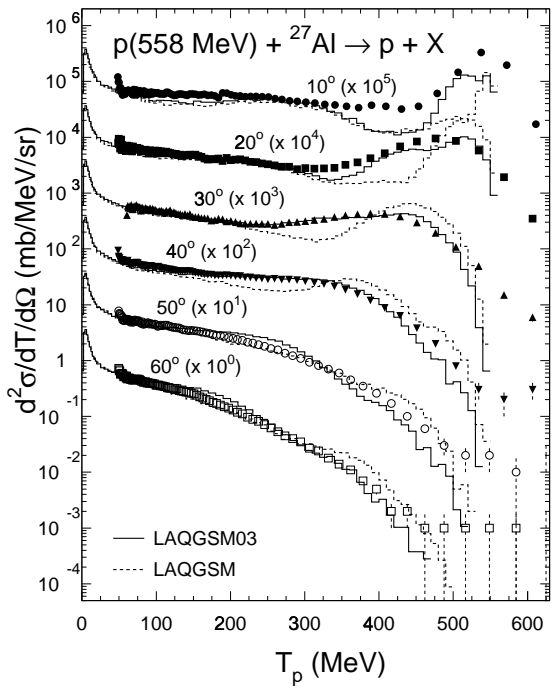

**FIGURE 3.** Proton spectra at 10°, 20°, 30°, 40°, 50°, and 60° from the reaction 558 MeV p + Al. Symbols are experimental data from [15], dashed and solid histograms are results from the old version of LAQGSM [2] and from LAQGSM03, respectively.

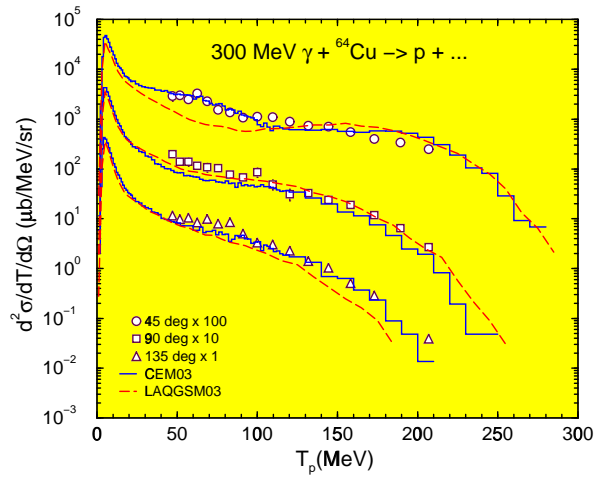

**FIGURE 4.** Proton spectra at 45°, 90°, and 135° from the reaction 300 MeV $\gamma$ + Cu. Symbols are experimental data from [16], histograms and dashed lines are results of CEM03 and LAQGSM03, respectively.

We have verified and compared these codes to a large variety of experimental particle spectra from *NN*, *NA*, $\gamma A$, and *AA* reactions at different energies and found that generally the new codes describe particle spectra much better than their precursors. Two examples of proton spectra, from 558 MeV p + Al and 300 MeV $\gamma$ + Cu are shown in Figs. 3 and 4, respectively. From Fig. 3, we see that LAQGSM03 describes much better the measured spectra than its precursor, especially at forward angles. Nevertheless, the agreement is not yet completely satisfactory and there is room for further improvements. We choose especially this particular reaction showing some problems to point out that although major improvements have been made in these codes in comparison with their precursors, they are far from being perfect and need further development, just as do all other INC-type models: to the best of our knowledge, no models or codes currently described in the literature can describe without problems spectra of forward-emitted nucleons from arbitrary reactions (see, *e.g.*, [6, 7, 18] and references therein).

After testing our codes against experimental particle spectra, we then compare them to various isotope-production cross sections from spallation, fission, and fragmentation reactions, that were described quite well by the previous versions of these codes [4, 5]. We find that usually the new codes describe such reactions as well as and often better than their precursors (see, *e.g.*, the first three top panels in Fig. 5). Nevertheless, our new codes fail to reproduce correctly production of fission-like heavy fragments from reactions with medium and light nuclear targets at intermediate energies (see the solid lines on the bottom two panels of Fig. 5), just as do their precursors and other currently available models (see, *e.g.*, [17, 19]). Such nuclear targets are considered too light to fission in conventional codes (including GEM2 and all codes currently employed in large-scale transport models). Similarly, the fragments are too light to be produced as spallation residues at these intermediate energies and too heavy to be produced via standard evaporation models.

To consider reactions involving very highly excited nuclei ($E^* > 2 - 3$ MeV/A) and as a possible way to solve the problem with emission of heavy fragments from medium-mass targets, we have incorporated into CEM03 and LAQGSM03 the statistical multifragmentation model code SMM by Botvina *et al.* [20], as a possible reaction mechanism occurring after the preequilibrium stage. We use here the default version (1997) of SMM kindly supplied us by Botvina, without any changes or fitting of parameters, except one: the only parameter of SMM we had to define in our combined codes CEM03+SMM and LAQGSM03+SMM was the value of the excitation energy of the excited compound nucleus produced after the preequilibrium stage of a reaction when we would consider to undergo multifragmentation processes as described by SMM instead of tra-ditional evaporation/fission mechanisms considered by standard CEM03 and LAQGSM03 for such reactions. We choose here the value of $E^* = 2$ MeV/nucleon for this transition: when the excitation of compound nuclei is $E^* \leq 2$ MeV/nucleon, we describe the reactions using the evaporation/fission model of GEM2 [3], standard in CEM03 and LAQGSM03, while at higher excitations, we use the SMM instead of GEM2.

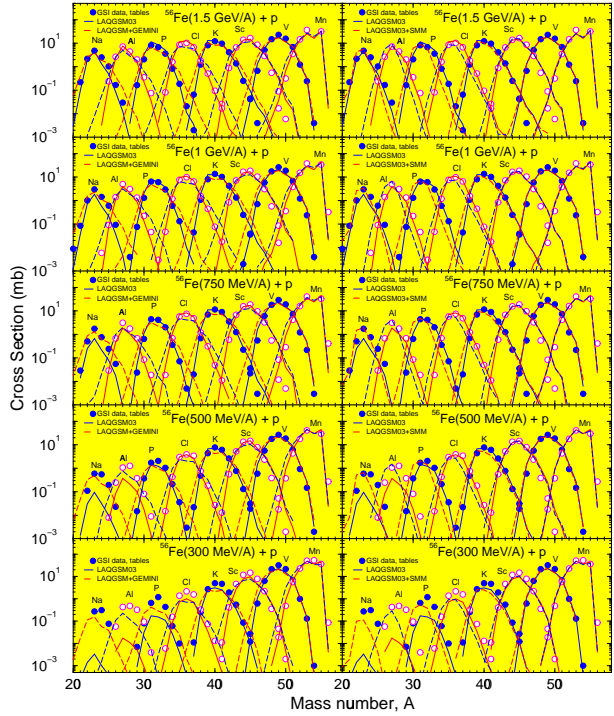

**FIGURE 5.** Experimental [17] mass distributions of the yields of eight isotopes from Na to Mn produced in the reactions 1500, 1000, 750, 500, and 300 MeV/A $^{56}$Fe + p compared with LAQGSM03 (solid lines on both panels), LAQGSM03+GEMINI (dashed lines, left panel), and LAQGSM03+SMM (dashed lines, right panel) results, respectively.

As we see from the right panels of Fig. 5, the combined LAQGSM03+SMM code reproduces quite well (dashed lines) the yields of all products, including the ones of heavy fragments from the reactions 300 and 500 MeV p + Fe shown in the two bottom panels, that can not be described by the standard LAQGSM03 (solid lines). Similar results were obtained by CEM03+SMM, and also for other reactions. This result does not mean, however, that the only possible mechanism for the production of heavy fragments from such intermediate-energy reactions on medium-mass nuclei is multifragmentation. The reactions shown in Fig. 5 (and other similar reactions) can be also described by versions of our codes when they are merged [19] with the fission-like sequential-binary-decay model GEMINI by Charity [21] (see the left panels of Fig. 5). Again, we used default parameters for GEMINI. This makes it more difficult to determine the mechanism of such processes. We think that for such intermediate-energy proton-induced reactions the contribution of multifragmentation to the production of heavy fragments should not be not very significant due to the relatively low excitation energies involved. Such fragments are more likely to be produced via the fission-like binary decays modeled by GEMINI. Multifragmentation processes are important and should be considered in reactions involving higher excitation energies; at excitations probably higher than the 2 MeV/nucleon considered here. We conclude that it is impossible to make a correct choice between fission-like and fragmentation reaction mechanisms involved in our $p+^{56}$Fe (or other similar) reactions merely by comparing model results with the measurements of only product cross sections; addressing this question will probably require analysis of two- or multi-particle correlation measurements.

To summarize, we have developed new versions of our codes, CEM03 and LAQGSM03, that describe quite well a variety of nuclear reactions at energies up to hundreds of GeV/nucleon better than their precursors. (Energies above about 5 GeV are only accessible to LAQGSM.) What is more, our codes provide reasonable results even for low-energy reactions, where they are not easy to justify from a fundamental-physics point of view. We show in Fig. 6 that CEM03 describes reasonably well even the experimental neutron spectra from the reaction 14.7 MeV n + $^{235}$U and gives results not far from those of the Los Alamos fission model by Madland and Nix [23] especially developed for such low-energy reactions. The role of the INC for such reactions is minor: the contribution to the total neutron spectrum as calculated by CEM03 from the INC is of the order of only one percent, as is the contribution from the preequilibrium processes, and the shapes of the INC and preequilibrium components of the spectrum are very similar. This means we could describe this part of the total spectrum either in terms of the INC or in terms of only preequilibrium emission, as is done by such codes as GNASH and TALYS. For CEM03, considering the INC as an initial stage of reactions even at such low energies is important, as INC provides the value of the number of excitons (excited nucleons and holes) as an input for the subsequent, preequilibrium stage of a reaction, and we do not need to treat this as an independent input parameter as is done in two-step preequilibrium plus evaporation/fission models. At the same time, the relative contribution to nucleon spectra and other characteristics of such low-energy nuclear reactions from the INC is very small, and should not affect significantly the final results, even though it is used at such low energies that its use can not be justified on physical grounds.

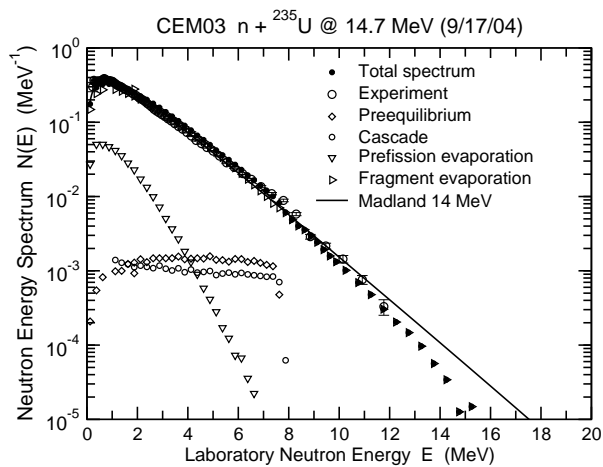

**FIGURE 6.** Comparison of the present CEM03 results with experimental data (big open circles) [22] and calculations by the Los Alamos fission model (line) [23] for the spectrum of neutrons from 14.7 MeV n + $^{235}$U. Contributions from the INC, preequilibrium, pre-fission evaporation, and evaporation from fission fragments to the total CEM03 results (filled circles) are shown by small open circles, diamonds, down triangles, and right triangles, respectively.

We thank Dr. Helder Duarte for providing us with numerical values of experimental cross sections from his collection, useful discussions, and help. We are grateful to Dr. Igor Pshenichnov for sending us the $\gamma p$ and $\gamma n$ event generators used in their Moscow photonuclear reaction INC and to Dr. Alexander Botvina for providing us with the 1997 version of his SMM code. This work was partially supported by the US Department of Energy, Moldovan-US Bilateral Grants Program, CRDF Project MP2-3045, and the NASA ATP01 Grant NRA-01-01-ATP-066.